# Spatial imaging of the H$_2^+$ vibrational wave function at the quantum limit

L. Ph. H. Schmidt*, T. Jahnke, A. Czasch, M. Schöffler, H. Schmidt-Böcking, and R. Dörner
Institut für Kernphysik, Goethe-Universität, Max-von-Laue-Strasse. 1, 60438 Frankfurt am Main, Germany

We experimentally obtained a direct image of the nuclear wave functions of H$_2^+$ by dissociating the molecule via electron attachment and determining the vibrational state using the COLTRIMS technique. Our experiment visualizes the nodal structure of different vibrational states. We compare our results to the widely used reflection approximation and to quantum simulations and discuss the limits of position measurements in molecules imposed by the uncertainty principle.

*Lothar.Schmidt@atom.uni-frankfurt.de

Even though quantum mechanics underlies almost every theoretical description of molecules, we often think of molecules as having well defined geometrical structures in real space. A very prominent case for that simplification inspired by classical physics can be found, e.g., in the common ball and stick models of molecules. The rational for this is based on the Born-Oppenheimer approximation. It assumes that the electronic and the nuclear part of a molecular wave function can be separated.

Imaging of the nuclear part of the molecular structure today is achieved mainly by x-ray [1] or electron [2,3] scattering on crystallized samples. Coulomb explosion imaging is a powerful tool to investigate the vibrational ground state of diatomic molecules [4] as well as vibrational wave packets [5]. Other techniques like diffraction of coherent x-ray burst from upcoming free electron lasers (FEL) [6,7], Coulomb explosion imaging of larger clusters [8], imaging by higher harmonic generation [9,10], or laser imaging by rescattering electrons [11] are currently developed. They are promising candidates to deliver structural images even of single molecules in the gas phase. However, vibrationally excited molecules have wave functions with complex structure. In particular, they have nodes in real space, i.e., positions at which the probability to find a nucleus is zero even though the molecule vibrates across these nodes. This fact is rather puzzling to our imagination guided by classical physics intuition where a particle cannot move from one to another position without passing all points in between. Already the Heisenberg uncertainty principle shows that there is a fundamental limit to our classical comprehension softening this paradox. The question arises of what the reality of the spatial structure of vibrational wave functions actually is and whether it can be actually observed in an experiment given the limits imposed by the uncertainty principle. We achieved this goal (Fig. 1) for the most fundamental molecule imaginable - the H$_2^+$ molecule.

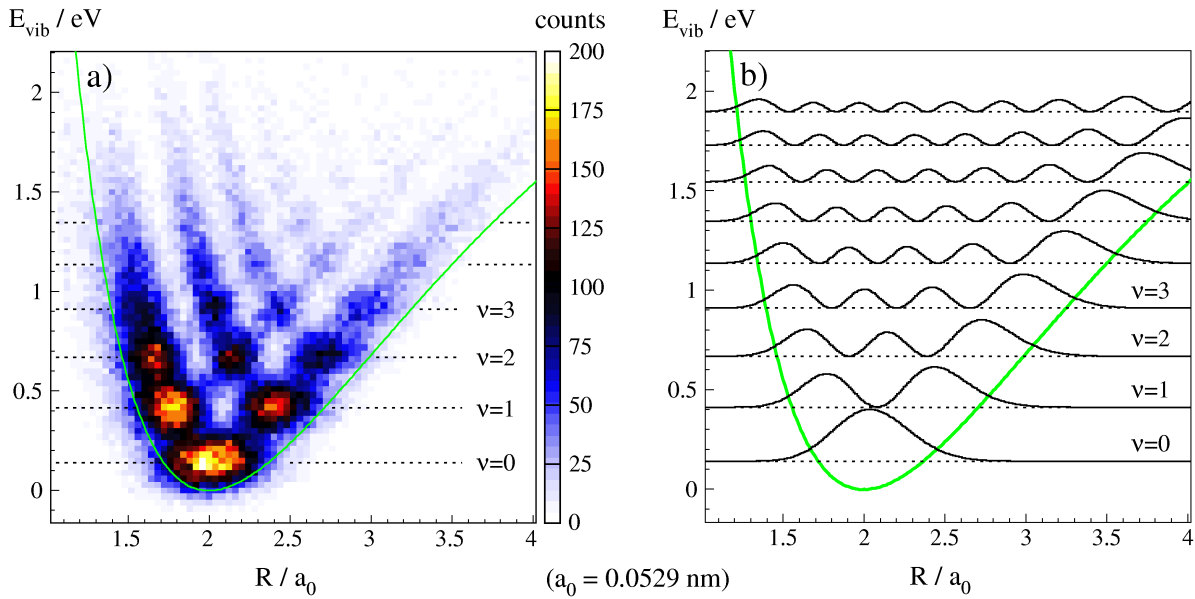

FIG 1. Image of the nuclear wave functions of H$_2^+$. a) Measured vibrational energy $E_{vib}$ versus internuclear distance $R$ calculated from the kinetic energy release of the dissociation as described below. The potential energy curve of H$_2^+$(1s$\sigma_g$) extracted from [12] is shown as green line. b) Theoretical spatial density of the nine lowest vibrational states numerically calculated using the Born-Oppenheimer approximation.

The $1s\sigma_g$ electronic ground state of $H_2^+$ has a potential minimum at 0.107 nm = 2.0 $a_0$ (Bohr radius $a_0$ = 1 a.u. = 0.0529 nm). The energy spacing between the ground and the first vibrationally excited state is 0.272 eV, which is spectroscopically well characterized. Spectroscopy in the energy domain, however, characterizes each state by only one number: it does not give a real space image of the wave function itself. To obtain an image of the vibrational wave function we need a highly precise measurement of internuclear distance of a single molecule.

One of the most powerful tools nowadays for a single molecule internuclear distance measurement is Coulomb explosion imaging [13]. Here the electronic wave function of the molecule under investigation is altered in order to promote the molecule to a steeply repulsive energy surface. The nuclei are then rapidly driven apart by Coulomb repulsion. As the repulsive potential is known, it maps an initial internuclear distance to a final state kinetic energy which can be measured (Fig. 2). Repeating this measurement many times on an ensemble of molecules yields a probability distribution of all occurring internuclear distances. This technique was pioneered employing fast ion beams, where molecular ions were stripped off their electrons by rapidly passing through a thin foil and then imaging the ionic fragments [13]. The concept was later ported to the time domain by removing electrons with an ultrashort laser pulse [14] and more recently it was also used in coincidence studies revealing correlations between electronic and nuclear degrees of freedom [4,15,16].

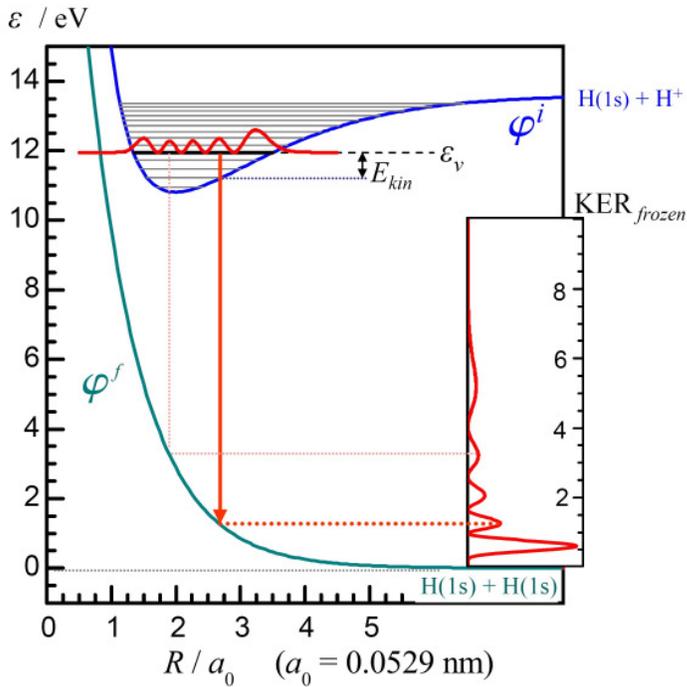

FIG. 2. Single molecule imaging using the reflection approximation for $H_2^+$. The molecule is promoted by a vertical electronic transition to the repulsive potential energy curve $b^3\Sigma_u^+$ (green line) which is described by the function $\varepsilon = \varphi^f(R)$. Within a few 10 fs it dissociates along this curve. After approximately 1500 ns we measure the kinetic energy release (KER) of the H(1s) fragments. In the standard frozen nuclei version of the reflection approximation the relationship between KER and the internuclear distance $R$ is given by $KER_{frozen} = \varphi^f(R)$. The resulting KER distribution is shown in the inset.

In the present Letter we take the principle of Coulomb explosion imaging to a new level of accuracy in order to image the wave function of $H_2^+$ molecules in different vibrational states and show how the uncertainty relationship influences the interpretation of the images obtained. The experimentally obtained internuclear distance $R$ shown in Fig. 1 and 3 are, as pointed out, not directly accessible. They are deduced from an energy measurement of the fragments as explained in more detail below. It turns out that it is this conversion where the quantum features of the measurement process enter.

The 2.5 keV $H_2^+$ ions investigated in our experiment are created in a penning ion source populating many excited states [17]. These vibrationally hot $H_2^+(1s\sigma_g)$ ions are neutralized by capturing an electron in collision with a small beam of cold He atoms produced by expanding gas at 16 bar, 140 K through a 30 μm nozzle. By detecting the times of flight and the emission angles of both molecular fragments in coincidence with the remaining $He^+$ target using cold target recoil ion momentum spectroscopy (ColTRIMS) [18, 19] we determine the vector momenta of all final state particles. This yields the kinetic energy release (KER) of the molecular breakup and the energy transferred from the motion of the projectile into internal degrees of freedom.

The neutral molecular fragments hit a 80 mm position sensitive micro channel plate detector with hexagonal delayline anode. This detector was optimized to detect two particles hitting the detector within a short time delay of only a few 10 ns. The KER was determined from the separation of the fragments in time and position. Beside the direct electron capture into the $b^3\Sigma_u^+$ state of $H_2$, which we used to map the internuclear distances, several higher excited states of the neutral molecule are populated. Predissociation of certain vibrational levels of the $c^3\Pi_u$ state results in a large number of peaks at the KER distribution starting at 7.286 eV [20]. This structure was used to calibrate the KER measurement.

Using ColTRIMS with an ion optics which minimizes the effect of the target size the $He^+$ momentum component parallel to the $H_2^+$ ion beam could be measured with resolution better than 0.04 a.u. (FWHM). From this parallel component of the recoil ion we obtain the energy gain of the fast projectile [21, 22]. The resolution is not only sufficient to resolve the final electronic states (which are separated by several eV) but also the initial vibrational excitation of the $H_2^+$. Thus we are able to distinguish different initial vibrational states in our measurement.

We select events where the electron is transferred from the helium atom into the $2p\sigma_u$ state of the hydrogen molecule leading to a rapid dissociation into two neutral hydrogen atoms. To map the $H_2^+$ nuclear wave function we used only those events with small transversal momentum (< 1.5 a.u.= $3 \cdot 10^{-24}$ kg m s$^{-1}$) from the measured data.

The principle of molecular explosion imaging is shown in Fig. 2. Initially $H_2^+(1s\sigma_g)$ is in a specific vibrational state ν. By attaching an electron it is promoted to a repulsive curve of $H_2$. It dissociates along this curve resulting in two H atoms in their electronic ground state with a KER which is measured. In the widely used frozen nuclei reflection approximation [23] the KER can be converted to the internuclear distance $R$ at the instant of the electronic transition by using the equation $KER_{frozen}= \varphi^f(R)$ where $\varphi^f$ describes the potential energy curve (PEC) of the dissociative state (we define $\varphi^f(\infty)=0$). Applying this to our data yields the green dots in Fig. 3a+b. These results are in good agreement for the ground state, but for the excited states the experimental minima and maxima are shifted significantly with respect to the calculated wave function (black line).

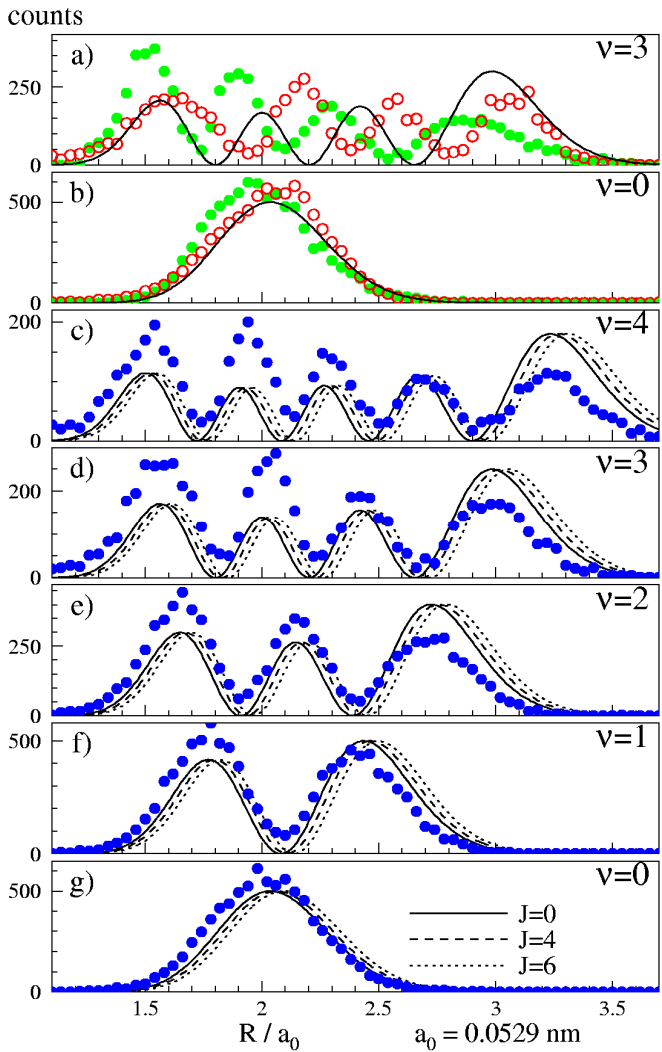

FIG. 3. Spatial density of the lower vibrational states of $H_2^+$ with vibrational quantum number ν=0 to 4. (a),(b) The internuclear distance R was calculated from the measured data either by the standard frozen nuclei reflection approximation (green dots) or the moving nuclei reflection approximation (red open circles). For both approximations the positions of the nodes do not coincide with the theoretical prediction (black solid line). (c)-(g) The mean value of both approximations (blue dots) is compared to the theoretical distributions of the rotational ground state (solid line) and rotationally excited states characterized by the quantum number $J$ which is not determined in the experiment.

What is the reason for this severe disagreement? The frozen nuclei reflection approximation implicitly assumes the nuclei to be at rest at the instant of the electronic transition – the nuclei start their promotion along the dissociative curve with zero velocity. On the other hand, the Born-Oppenheimer approximation suggests that the nuclear wave function remains unchanged upon an electronic transition. In classical terms this implies that not only the positions of the nuclei are maintained (vertical "Franck-Condon" transition), but also the momenta, i.e. the kinetic energy $E_{kin}$ of the nuclei is preserved upon electronic excitation. This can be incorporated in a generalized version of the reflection approximation involving moving nuclei. Thus for each $R$ the $KER_{moving}$ is the sum of the KER accumulated along the final state PEC plus $E_{kin}$ determined by the energy of the vibrational state $\varepsilon_v$ with respect to the initial state PEC at the distance $R$. For higher vibrational states the results (red open circles in Fig. 3a) are significantly different to the frozen nuclei reflection approximation (green dots). It does, however, clearly overestimate the effect of nuclear motion. Heuristically we therefore use for each event the mean value from both models. This assumes that approximately half of the kinetic energy $E_{kin}$ at each $R$ is converted to KER. The resulting blue dots in Fig. 3c-g show a good agreement with the position of the nodes from the theoretical $R$ distribution but the theoretical peak height is not well reproduced. While we image the nuclear wave function only in case of an electron transfer into the $2p\sigma_u$ state, we observe a convolution of the initial state with the probability of the electronic transition. From our measurement we conclude that the electron transition probability changes by a factor 2 within the Franck-Condon region of vibrational excited states with the higher probability at small internuclear distances. Nevertheless, compared to alternative electronic transitions initializing dissociation, such as radiative transitions [24], this probability only slightly interferes the imaging of the nuclear wave function.

We note that both versions of the reflection approximation violate the Heisenberg uncertainty relation. Therefore it is surprising that a simple averaging over the two extreme scenarios yields good agreement between the calculated and measured location of the nodes of the vibrational wave functions (Fig. 3(c)-(g)).

To gain further insight into the quantum aspects of the mapping of position space to KER we examined the quantum version of the reflection approximation by calculating the overlap integral of the initial state and final states for fixed KERs. The resulting KER distributions are in good agreement with the measured positions of the peaks (Fig. 3(b) and (c)). An inspection of the overlap visualized in Fig. 4(b) shows how $R$ is mapped to KER. For none of the KER the main contribution to the overlap integral (light colored areas) comes from the position suggested by the reflection approximation which is shown by the green line. The smallest KER exclusively arises from larger $R$ but the high KER samples the whole Franck-Condon region. Given this broad sampling of space for a specific KER it might be surprising why nodes appear in the measured and calculated KER. Clearly these nodes in the KER distributions do not correspond to a particular $R$. They are the result of cancellation of positive (fig 4b, red areas) and negative (blue areas) contributions to the overlap between initial and final state.

In conclusion our study provides a textbook image of the vibrational states of $H_2^+$ within the limits given to spatial imaging by quantum mechanics. For alternative single molecule imaging scenarios like superintense x-ray scattering at future FELs similar limits will also hold. Conceptually interesting questions can be raised once two measurement techniques are employed simultaneously on the same molecule. The answers of such combined imaging experiments should depend on the sequence of the measurements, since quantum mechanically the measurement itself alters the state by projecting it onto one of the eigenstates of the measurement apparatus.

The experimental work was supported by the Deutsche Forschungsgemeinschaft.

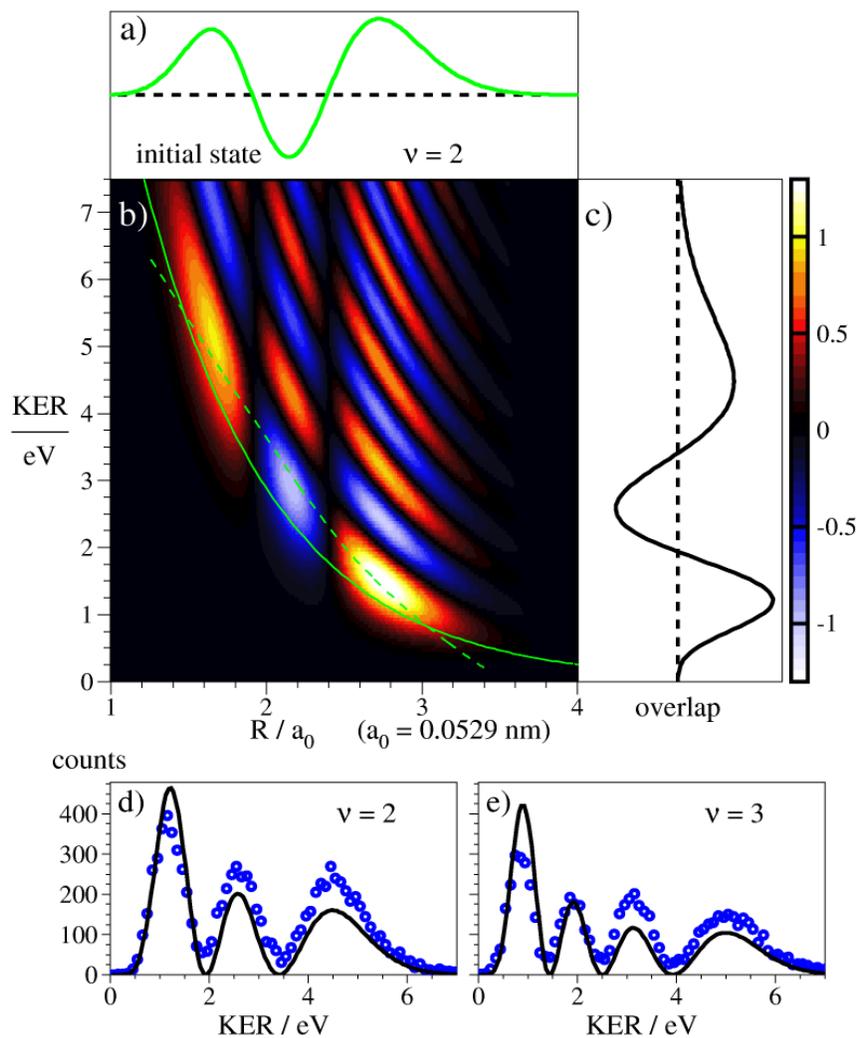

FIG. 4. Quantum mechanical calculation of the KER distribution. a) Initial state wave function of $H_2^+$ $v = 2$. b) Product of initial state and the continuum states of $H_2$ ($b^3\Sigma_u^+$) parametrized by KER. The frozen nuclei and moving nuclei reflection approximation constitute a strong correlation between $R$ an KER as depicted by the green solid and green dashed lines. Integrating this product over $R$ yields the KER dependent overlap shown in panel c). The square of the overlap is compared to the experimental results in d) and e) shows the same for the vibrational state $v=3$.